\documentclass{elsart}
 \usepackage{graphicx}
 \usepackage{epsfig}
\usepackage{amssymb}
\usepackage[latin1]{inputenc}
\usepackage{graphicx}  
\usepackage{bm}        
\usepackage{latexsym}
\usepackage{mathrsfs}
\usepackage{amssymb}
\usepackage{amsmath}
\usepackage{amscd}
\usepackage{color}
\usepackage{verbatim}

\journal{Physica D}

\begin{document}

\begin{frontmatter}

\title{Quantum energy decays and decoherence in discrete baths}

\author[PR,AM]{M.~D.~Galiceanu},
\author[PR]{M.~W.~Beims\corauthref{cor}}, and
\corauth[cor]{Corresponding author} \ead{mbeims@fisica.ufpr.br;
{\it Tel.}: +55 41 3361 3349; {\it Fax}: +55 41 3361 3418}
\author[DR]{W.~Strunz}

\address[PR]{Departamento de F\'{\i}sica,
             Universidade Federal do Paran\'a,\\
             81531-990 Curitiba, Brazil}

\address[AM]{Departamento de F\'{\i}sica,
             Universidade Federal do Amazonas,\\
             69077-000 Manaus, Brazil}

\address[DR]{Technische Universit\"at Dresden, 
	     D-01062 Dresden, Germany}
\maketitle

\begin{abstract}
The quantum average energy decay and the purity decay are studied for a 
system particle as a function of the number of constituents of a discrete 
bath model. The system particle is subjected to two distinct physical
situations: the harmonic oscillator (HO) and the Morse potential. The 
environment (bath) is composed by a {\it finite} number $N$ of uncoupled 
HOs, characterizing the structured bath, which in the limit 
$N\to\infty$ is assumed to have an ohmic, sub-ohmic or super-ohmic 
spectral density. For very low values of $N$ the mean energy and purity 
remain constant in time but starts to decay for intermediate values 
($10\lesssim N\lesssim 20$), where two distinct time regimes are observed: 
two exponential decays for relatively short times and a power-law decay 
for larger times. In this interval of $N$ decoherence occurs for short 
times and a non-Markovian dynamics is expected for larger times. When $N$ 
increases, energy and coherence decay very fast and a Markovian dynamics 
is expected to occur. Wave packet dynamics is used to determine the 
evolution of the particle inside the system potentials.

\end{abstract}
\end{frontmatter}

\section{Introduction}\label{intro}
The quantum dissipation and decoherence are analyzed here for an open 
system interacting with its environment by collision processes. The problem 
(System+Environment+In\-te\-rac\-tion) is conservative 
but, due to energy exchange between system and environment, the
{\it system} can be seen as an open system with dissipation.
Such theoretical model has been proposed and studied in quantum 
\cite{r9,r14,r15} and classical \cite{r14,r15,r13} systems. Various models
were developed to treat such open system.
We start by mention the methods 
which have focused on an explicit quantum dynamical treatment of the 
system+bath dynamics: the multiconfiguration time-dependent Hartree (MCTDH) 
technique for wave packet propagation \cite{r_wang}, the Gaussian-MCTDH 
approach \cite{r_burg1}, the effective-mode representation\cite{r_burg2}, 
and the local coherent-state approximation to
system-bath dynamics 
\cite{r_martinazzo}. Another way of studying
the system+bath dynamics is 
to solve the non-Markovian master equations \cite{r_meier,r_ishizaki}, 
including some semiclassical approaches \cite{r_stock1,r_stock2,r_frank1}.

Here a non-Markovian quantum trajectory theory is used, named 
non-Mar\-ko\-vian quantum-state diffusion, that describes the dynamics
of a quantum system coupled linearly via position to a environment
\cite{r11,r12,r3,r5}. Differently from most the previous studies, the 
environment is composed by a {\it finite} number, $N$, of uncoupled HOs, 
as studied before for classical systems \cite{r8,r1}. In such 
cases we say to have a discrete or structured bath. Structured baths have 
become important since they describe realistic physical situations of 
non-equilibrium physics. One major recent application is the energy transfer 
between a light-harvesting protein and a reaction center protein
\cite{r_sarovar}. Understanding the physics encountered in this 
process will help exploring its huge potential \cite{r_scholes}. Some 
experiments in which the non-Markovian behavior arises due to the 
discreteness of the bath can be mentioned: high-Q microwave cavities, 
quantum optics in materials with a photonic band gap, output coupling 
from a Bose-Einstein condensate to create an atom laser 
\cite{r16,r17,r18,r19,r20} or the decoherence phenomenon 
\cite{r_dec1,r_dec2,r_dec3,r_wong}. { From the classical point 
of view, in the context of finite baths with uncoupled HO, there are some 
works which analyze the effect of discrete (structured) baths on the system 
energy decay \cite{r13,r8,r1}. The main result found is that finite baths 
may induce a 
non-Markovian dynamics on the system particle. From the quantum point of view, 
most works \cite{r13,r_burg1,r_martinazzo,r_frank1,r_wong,r_koch} focus on 
the analysis of changing the frequencies distribution and the coupling 
strength between system and bath.

We analyze systematically the effect of increasing the number
of oscillators $N$ from the bath on the system energy decay.
In most cases we vary $N=1\to 100$.
As $N$ increases, the quantum simulations, usually done here for $500$ 
realizations of the bath, need very long computational times.
For very low values of $N$ the mean time energy decay keeps constant but 
starts to decay for intermediate values of $N$. For $10\le N\le 50$ a 
transition to exponential decay for shorter times and a power-law decay 
for larger times is observed. For higher values $N>100$ only exponential 
decay is observed. In the interval $20\le N\le 50$ an additional 
power-law decay between the two previously mentioned behaviors is observed.}

The paper is organized as follows. In Section \ref{linear_q} we introduce 
the basic concepts of non-Markovian quantum-state diffusion and derive a 
new simplified stochastic Schr\"{o}dinger equation valid for {\it finite}
baths. In Section \ref{HO} the simplified equation is applied to analyze
the $N$ dependence of some useful physical quantities such as the energy 
decay, the average position, and the purity (decoherence). In this section 
harmonic potential is considered. In Section \ref{morse} a Morse potential
is used for the system particle. In the last section we present a summary 
and the conclusions.

\section{Linear non-Markovian quantum-state diffusion}\label{linear_q}

The non-Markovian quantum-state diffusion (QSD) equation is based on a 
standard model of open system dynamics: a quantum system interacting with 
a bosonic environment with the total Hamiltonian

\begin{equation}\label{H_tot}  
H_{tot}=H+ \sum_{\lambda} g_{\lambda}(L a^{+}_{\lambda}+L^{+} a_{\lambda})+ 
\hbar \sum_{\lambda} \omega_{\lambda} a^{+}_{\lambda} a_{\lambda},
\end{equation}
where $H$ is the Hamiltonian of the system of interest, $L$ is a
system operator coupling to environment and $a^{+}_{\lambda}$, 
$a_{\lambda}$ are the raising and lowering operators, with the
property 
$[a_{\lambda},a^{+}_{\lambda ^{'}}]=\delta_{\lambda
\lambda^{'}}$. The linear 
non-Markovian QSD equation
\cite{r3,r5,r2,r4,r6} is given by

\begin{equation}\label{sch_eq}  \hbar  \left| \dot{\psi}_t \right> =-iH' \left| \psi_t \right> 
+ L z^{*}_{t} \left| \psi_t \right> -L^{+} \int^{t}_{0} ds K(t-s)
 \frac{\delta \left|\psi_t\right>}{\delta z^{*}_{s}}, \end{equation}

where was assumed a factorized total initial state 
$\left|\psi_0 \right> \left|0_1\right> \left|0_2\right>
\cdots \left|0_{\lambda}\right>\cdots$, with an arbitrary system state
$\left| \psi_0 \right>$ and all environmental oscillators in their
ground state $\left|0_{\lambda}\right>$. 

\begin{equation}\label{z_t} 
 z^{*}_{t}=-i \sum_{\lambda} g^{*}_{\lambda} z^{*}_{\lambda} e^{i \omega_{\lambda} t}
\end{equation}
is a complex Gaussian process with zero mean and correlations 
$M[z^{*}_{t} z_{s}]= K(t-s)=\sum_{\lambda} |g_{\lambda}|^2 e^{-i
\omega_{\lambda} (t-s)}$, 
where $M[ ]$ denotes the {\it ensemble
average} over the classical driving 
noise. The form of $K(t-s)$
corresponds to the zero-temperature limit. 
$\lambda$ is the number
of oscillators from the bath, $g_{\lambda}$ is the 
coupling to
oscillator $\lambda$, $\omega_{\lambda}$ is its frequency and 
$z^{*}_{\lambda}=(x_{\lambda}+i y_{\lambda})/ \sqrt{2}$, where 
$(x_{\lambda},y_{\lambda})$ are normal (Gaussian) distributed real number with 
zero mean and deviation one. The many realizations to solve the stochastic 
Schr\"{o}dinger equation are done over these
normal distributed numbers. 
The functional derivative of equation
(\ref{sch_eq}) can be written in the 
first order approximation as a linear function of $\left|\psi_t\right>$

\begin{equation}\label{f_der}
 \frac{\delta \left|\psi_t\right>}{\delta z^{*}_{s}}=O(t,s,z^*) 
\left|\psi_t\right> = e^{-iH(t-s)} L e^{iH(t-s)} \left|\psi_t\right>.\end{equation}
Now, taking $L=L^{+}=q$, which is the position operator from the 
system, the stochastic Schr\"odinger equation becomes

\begin{equation}\label{lin_scheq}
  \hbar  \left| \dot{\psi}_t \right> =-iH' \left| \psi_t \right> +qz^{*}_{t} 
\left| \psi_t \right> -q \int^{t}_{0} ds K(t-s) e^{-iH(t-s)} q e^{iH(t-s)} 
\left| \psi_t \right>.
\end{equation}
In equation (\ref{sch_eq}) and (\ref{lin_scheq}), the Hamiltonian 
$H'=H(q,p) + q^2 A(t)$, with 
$A(t)=\sum_{\lambda}\frac{|g_{\lambda}|^2}{\omega_{\lambda}} 
(\cos \omega_{\lambda} t- 1)$, contains an additional potential 
term that turns out to be counterbalanced by a similar term 
arising from the memory integral \cite{r4}. Our approach is 
consistent with the Redfield theory \cite{Redfield92} in the 
long-time limit when $\int_0^t$ is replaced by $\int_0^{\infty}$.

Using $\left| \psi_t \right>=\sum_{n} c_n (t) \left| \phi_n \right>$, $H \left| \phi_n \right>=
\epsilon_n \left| \phi_n \right>$, $q=\sum_{n,m} q_{nm} \left| \phi_n \right> < \phi_m
|$, and orthonormal condition $<\phi_n | \phi_{n'}>=\delta_{n
n'}$, equation (\ref{lin_scheq}) transforms to

\begin{eqnarray}
\label{lin_coef}
 \hbar  \dot{c}_n (t)  &=& -i \epsilon_n c_n(t) -i A(t) \sum_{mm'} q_{nm}
  q_{mm'} c_{m'} (t) +z^{*}_{t} \sum_{m} q_{nm} c_m(t) \cr
& & \cr 
& &-\sum_{m,m'} q_{nm'} \bar {O}_{mm'}(t) c_{m'}(t),
\end{eqnarray}
where

\begin{equation}
\label{o_m}  \bar {O}_{mm'}(t)= <\phi_m | \bar {O} (t) \left| \phi_{m'}\right> 
= q_{mm'} \sum_{\lambda} |g_{\lambda}|^2 \frac{e^{-i [\omega_{\lambda}+(\epsilon_{m} - \epsilon_{m'})] t}-1}{-i[\omega_{\lambda}+(\epsilon_{m} - \epsilon_{m'})]},
\end{equation}
and

\begin{equation}
\label{o_t}  \bar {O} (t)=\int^{t}_{0} ds K(t-s) e^{-iH(t-s)} q e^{iH(t-s)}.
\end{equation}
For simplicity, we choose the Planck constant $\hbar=1$ 
throughout the paper. At next we analyze the following quantities:
{The quantum average of the {\it system} energy which is given by

\begin{equation}
\label{energy}
 \left<E\right>=\left<\psi_t|H|\psi_t\right>=\sum_n c_n^*(t) c_n(t) \epsilon_n,
\end{equation}
and the quantum average of the {\it system} position 

\begin{equation}\label{position}
 \left<q\right>_t=\left<\psi_t\right|q\left|\psi_t\right>=\sum_{n,m} c_n^*(t) c_m(t) q_{nm},
\end{equation}
and finally the purity which gives informations about the 
decoherence occurred in the
system \cite{r_frank1}:

\begin{equation}\label{purity}
P=\mbox{Tr}_c \left[\rho_s^2(t)\right]= \sum_{nm} M[c_nc_m^*]  M[c_mc_n^*],
\end{equation}
where $\rho_s(t) = M\left[\left|\psi_t\right>\left<\psi_t\right|\right]$ 
is the reduced density matrix.}

Now the stochastic Schr\"{o}dinger equation coupled to a finite bath, 
Eq.~(\ref{lin_coef}), will be implemented to calculate the energy decay,
{mean position and purity} for the system particle inside two 
types of potentials, widely used in physics: the harmonic potential and the 
Morse potential. We also consider the time evolution of the quantum average 
position.

\section{Harmonic potential coupled to $N$ oscillators}\label{HO}

{
In this section we consider a particle inside a harmonic potential
coupled to a discrete bath of $N$ HOs. This problem has exact 
solutions \cite{r14} which can be used to compare some quantities 
calculated with the approach used here. The frequencies 
$\omega_{\lambda}$ are distributed in the 
interval $(1.1,2.1)$ and are picked randomly from a frequencies 
generator, but once picked they will be kept fixed during the 
simulations, even for different bath realizations. 
In this approach, due to the finite number of HOs in the 
bath, the spectral density is determined numerically and is always 
structured for low values of $N$ (see \cite{r8} for more details in the 
classical case). The properties of the bath depend on the frequency's 
distribution of the $N$ HOs \cite{r8}. In the limit $N\to\infty$ we 
assume to have three distinct continuous distributions:  ohmic bath 
(quadratic), sub-ohmic and super-ohmic. Each oscillator will have a 
distinct $(x_{\lambda},y_{\lambda})$ pair, randomly chosen numbers with 
zero mean and deviation one. For each coefficient $c_n$ of 
Eq.~(\ref{lin_coef}) we consider a rectangular initial conditions: 
$c_n(0)=1+0 \cdot i$, with $n=1, \dots , n_{max}$, where $n_{max}$ is the 
last energy level considered. The first $n_{max}=15$ energy's levels from 
the harmonic potential are considered, with energy 
$E_n=\hbar \omega (n-\frac{1}{2})$. Using $\omega=\hbar=1.0$
the ground state energy level equals $0.5$. Thus our initial state
is given by

\begin{equation}
\left|\psi_0\right>=\frac{1}{\sqrt{n_{max}}}\sum_{n=1}^{n_{max}=15} 
\left|\phi_n\right>,
\label{phi0}
\end{equation}
which is an entanglement of all $15$ system levels.
In order to obtain the energy 
time decay and the average position as a function of the number of 
oscillators in the bath, Eq.~(\ref{lin_coef}) is integrated by a 
fourth-order Runge-Kutta integrator \cite{r_nr} and the values of 
$c_n(t)$ are introduced in Eqs.~(\ref{energy}) and (\ref{position}). 

In the first part of this section the frequencies of the harmonic 
oscillators from the bath 
will follow a quadratic distribution, making the spectral density to be 
linear $J(\omega\le \omega_{cut}) \propto \omega $ when  $N\to\infty$.
$500$ initial conditions for the pairs $(x_{\lambda},y_{\lambda})$ will be 
considered for all the studied cases. To render the comparison easier the 
coupling strength will be kept constant, namely $g_{\lambda}=0.01$.}

\begin{figure}[ht]
   \begin{center}
  \includegraphics[clip=,angle=0,width=9cm]{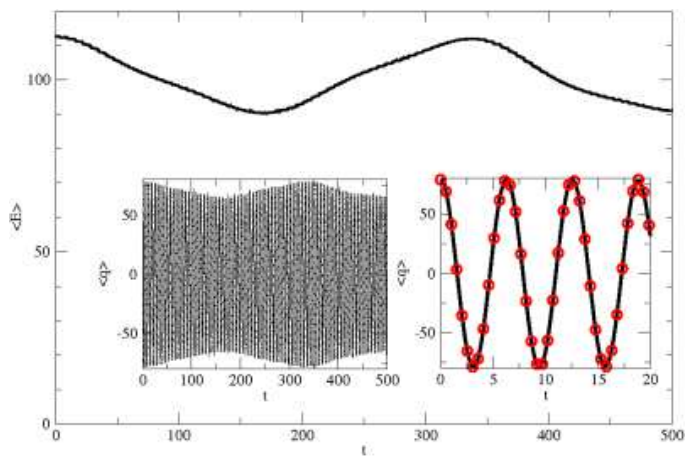}
  \caption{The average energy of a system, Eq. (\ref{energy}), coupled to a
   bath of one HO, with the coupling strength $g=0.01$. {The two 
  insets show the corresponding average positions, 
  Eq.(\ref{position}), for larger (left) and smaller (right) times. For 
  comparison the circles (right) show exact results obtained from 
  \cite{r14}}.} 
 \label{HO_1}
 \end{center}
\end{figure}

In Figure \ref{HO_1} the time evolution of the average energy $\left<E\right>$,
Eq.~(\ref{energy}), is plotted for a bath containing one harmonic 
oscillator with the frequency $\omega=2.09$. As inset graph is displayed 
the average position
$\left<q\right>$ from Eq.~(\ref{position}). One can clearly see that for times 
$t_d\sim 170$ the system loses around $20\%$ of its initial energy for the 
bath, but is able to regain it at a later time $t_R\sim 320$. This exchange 
of energy between system and environment will repeat itself, with frequency 
around $\omega\sim0.037$, for all integrated times (we checked it 
until $t=5000$). Thus, the mean energy over the return times (where the 
energy transferred into the bath returns to the system) is almost constant.
\begin{figure}[ht]
 \begin{center}
  \includegraphics[clip=,angle=0,width=9cm]{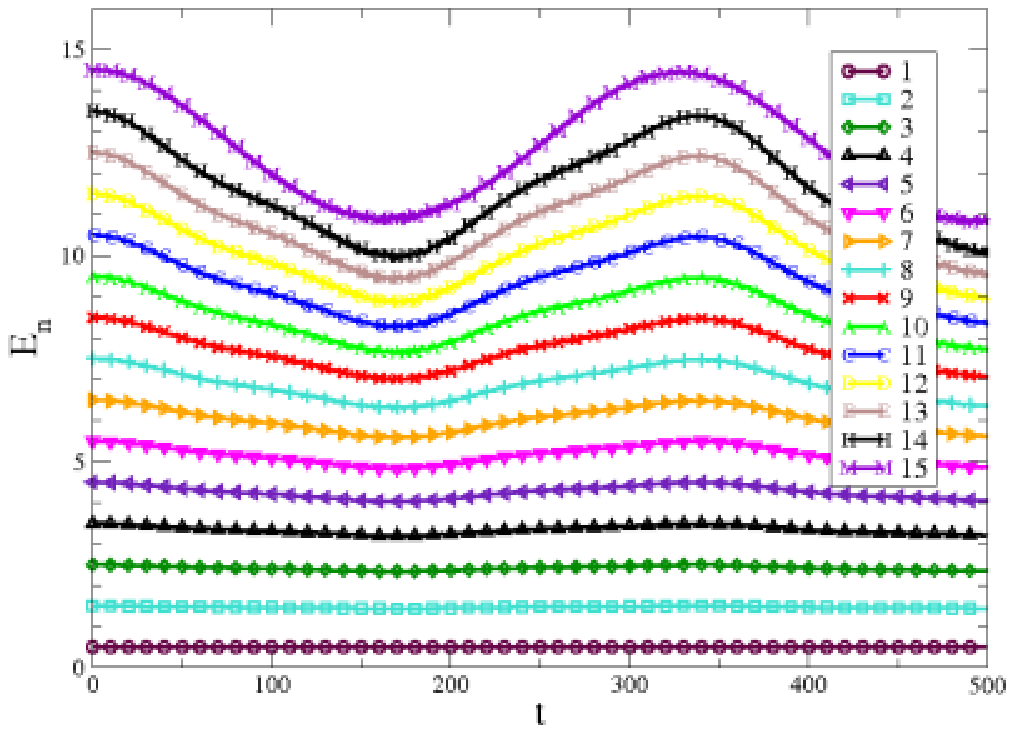}
  \caption{(Color online) The time evolution of all $15$ energy levels   
$E_n|c_n(t)|^2$} of a system coupled to a bath of one HO.
  \label{en_lev1} 
\end{center}
\end{figure}
The return times of the energy are strongly dependent on the frequency of the 
bath oscillator but independent of the coupling
intensity. This was checked 
for many coupling intensities. The
time behavior of the energy was also 
confirmed by the time
evolution of the average position: the system 
symmetrically
oscillates very fast around the center $\left<q\right>=0$, 
with an amplitude
which follows the pattern of energy decay. Namely, the 
minimum of
the energy corresponds to the minimum of the highest amplitude.
Figure \ref{en_lev1} shows the time evolution of all the $15$ individuals 
energy levels  {$E_n|c_n(t)|^2$}
of the harmonic potential for the same bath used in 
Fig.~\ref{HO_1}. Numbers denote the energy quantum number $n$. One can 
clearly notice that while the lowest states remain practically unchanged,
the higher energy levels oscillate following the energy exchange observed 
in Fig.~\ref{HO_1}. Thus only the highest states are responsible for the 
energy exchange.

\begin{figure}[ht]
 \begin{center}
 \includegraphics[clip=,angle=0,width=9cm]{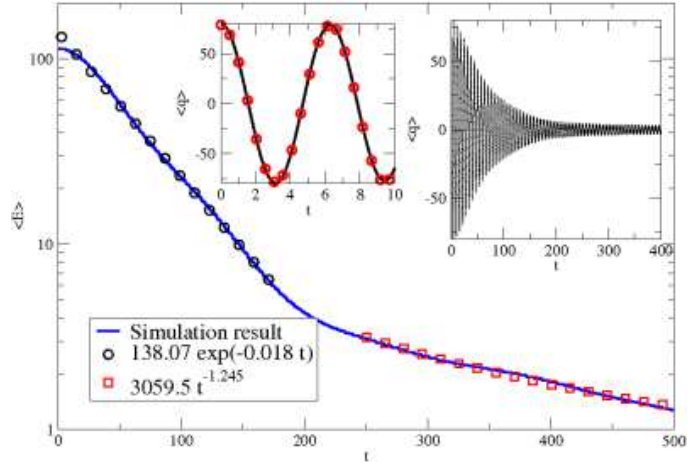} 
 \caption{(Color online) The average energy decay, Eq.~(\ref{energy}), 
for a bath of $10$ HOs, with the coupling strength equals to $0.01$. By 
 symbols we display the best fits. {The two insets show the corresponding 
average positions,  Eq.(\ref{position}), for larger (right) and smaller 
(left) times. For comparison the circles in the left inset show exact 
results obtained from  \cite{r14}}.}
  \label{HO_10} 
\end{center}
\end{figure}

In Figure \ref{HO_10} the energy decay is shown in a semi-log plot for a bath
of $N=10$ HOs. In the time limit studied here ($t_{max}=500$)
one can clearly see that the system energy will not be regained so that 
the time average from the system energy is not constant anymore.
However, we expect that for some later times $t>t_{max}$ the energy will return
to the system, since it is a finite system. The  qualitative behavior of the 
energy decay rate changes for different time intervals.
We show by symbols the best fits of the energy decay, namely an exponential 
fit, $\alpha=0.018$, for short times where a Markovian dynamics is expected 
and a power-law  fit, $\beta=1.245$, for large times, where a non-Markovian 
dynamics occurs. We say that the qualitative change from exponential to 
power-law decay occurs close to times $t_{\alpha\to\beta}$. As inset graph we 
display the corresponding average position.

\begin{figure}[ht]
 \begin{center}
  \includegraphics[clip=,angle=0,width=9cm]{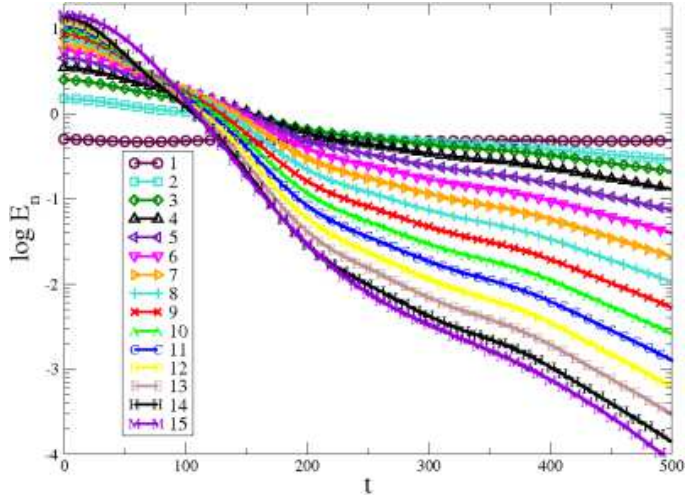}
  \caption{The time evolution of all $15$ energy levels 
{$E_n|c_n(t)|^2$}
of a system coupled to a bath containing $N=10$ HOs.}
  \label{en_lev10}
 \end{center}
\end{figure}

The time evolution of all the $15$ energy levels $E_n$ of a system coupled to 
a bath which has $N=10$ HOs is displayed in the semi-log plot
of Fig.~\ref{en_lev10}. Again it is clear that the higher energy levels decay
faster. Something similar was shown to occur with the decoherence rate which 
is correlated to the highest occupied states \cite{brumer04}. In the long 
times region the energy levels are totally inverted: the highest energy level 
becomes the smallest one while the ground state energy becomes the most 
predominant one, followed by the second energy level and so on. By comparison 
with Fig.~\ref{HO_10} we observe that the energy levels start to be totally
inverted for times close to $250$, i.~e.~$t_{\alpha\to\beta}$ where the transition
from exponential to power-law decay occurs. At these times it
can be observed in Fig.~\ref{en_lev10} that the decay ratio $dE_n/dt$ 
decreases, meaning that the 
energy inserted in the bath starts to influence back the system dynamics, 
generating the non-Markovian behavior and the observed power law decay.

\begin{figure}[ht] 
\begin{center} 
 \includegraphics[clip=,angle=0,width=9cm]{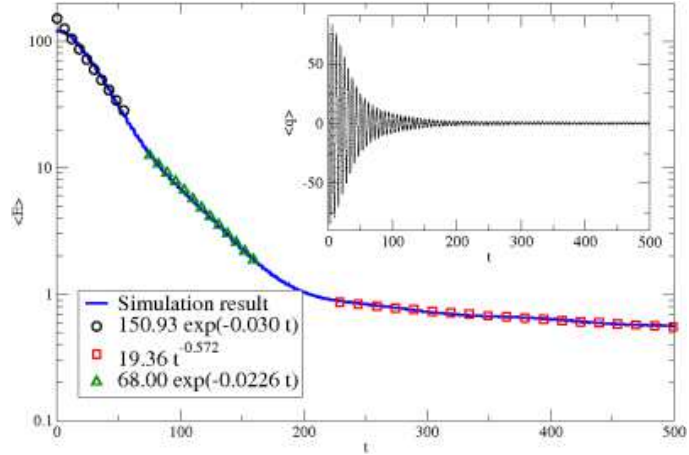}
  \caption{The average energy and its best fits for a bath of $N=20$ 
  harmonic oscillators. As inset we shown the average position.}
\label{HO_20}
 \end{center}
\end{figure}

The energy decay and the average position for a bath of $N=20$ harmonic 
oscillators is displayed in Fig.~ \ref{HO_20}. The system continues to 
behave dissipatively for the integrated time, with an exponential decay, 
$\left<E\right> \propto \exp(-0.030t)$ for $0<t\lesssim 50$ and a power-law 
decay  $\left<E\right> \propto t^{-0.572}$ for much larger times $t\gtrsim230$.
For an intermediate time region, starting at $t \sim 75$ and ending at 
$t\sim 190$, an additional exponential decay, 
$\left<E\right> \propto \exp(-0.0226t)$, was observed. Such kind of 
behaviors with two exponential decays was encountered until 
$N \sim 50$ and completely disappeared for $N\ge 100$. By keeping the same 
line of arguments of Fig.~\ref{en_lev10} we can say that, for a bath of 
$20 \le N \le 50$, the total inversion of the energy levels in times induces
a power-law decay and the non-Markovian behavior.

\begin{figure}[ht]
 \begin{center}
  \includegraphics[clip=,angle=0,width=9cm]{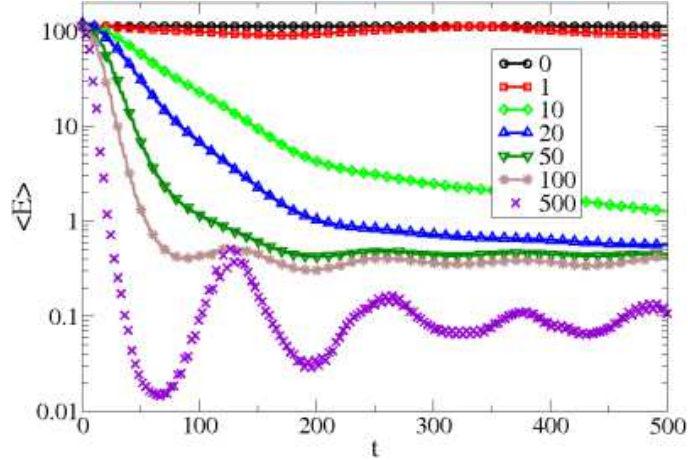}
  \caption{The energy decay for a bath with $N=1,10,20,50,100,500$ HOs. 
For $N=10,20,50$ we observe exponential and power-law decays.}
  \label{endecay_ho}
\end{center}
\end{figure}

\begin{figure}[ht]
 \begin{center}
  \includegraphics[clip=,angle=0,width=9cm]{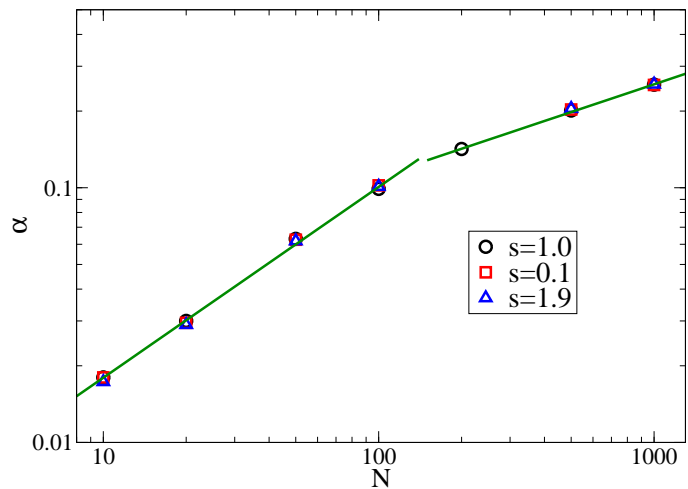}
  \caption{Exponential fits, $\alpha$, of the energy decay in the low
   times region for ohmic, subohmic, and superohmic baths as a function
  of $N$.}
  \label{expfits_ho}
 \end{center}
\end{figure}

\begin{figure}[ht]
 \begin{center}
  \includegraphics[clip=,angle=0,width=9cm]{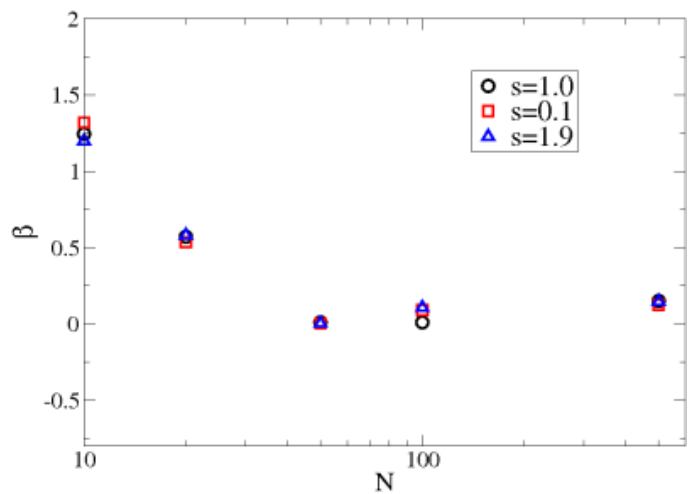}
  \caption{Power-law fits, $\beta$, of the energy decay in the long times
 region for ohmic, subohmic, and superohmic baths as a function of $N$.}
  \label{powerlawfits_ho}
 \end{center}
\end{figure}

In Figure \ref{endecay_ho} we summarize results of the energy decay for 
$N=1,10,20,50,100,$ $500$ HOs. The first observation is 
that the energy decays faster for larger $N$. For values $N=10,20,50$ we
observe an exponential decay for shorter times and power-law decay for 
larger times. Details of the decay exponents will be discussed later in
Figs.~\ref{expfits_ho} and \ref{powerlawfits_ho}. As $N$ increases the time
interval of power-law decay decreases and vanishes totally for $N=100$. 
Therefore, a non-Markovian motion is expected only for intermediate values 
of $N$. For $N\gtrsim 50$ a fluctuating behavior is observed and will be
explained below.

{ Instead of looking at the dissipation, it is also possible to 
analyze the purity from Eq.~(\ref{purity}) which gives informations about 
the decoherence occurred in the system.}
For our initial entangled state from Eq.~(\ref{phi0}),
the initial reduced density matrix $\rho_s(0)$ has $15$ diagonal terms and 
$210$ off-diagonal terms. The off-diagonal terms are characteristic of the 
quantum coherence. Total decoherence of the initial state occurs 
at times $t_{dec}$ when all off-diagonal terms vanish. For these times a
statistical mixture of the $15$ diagonal terms is reached and the 
the purity of the reduced density matrix is $P(t_{dec}) = 1/15\sim0.067$.
In this way it is possible to compare total decoherence times  $t_{dec}$ 
with dissipation times.

\begin{figure}[ht]
 \begin{center}
  \includegraphics[clip=,angle=0,width=9cm]{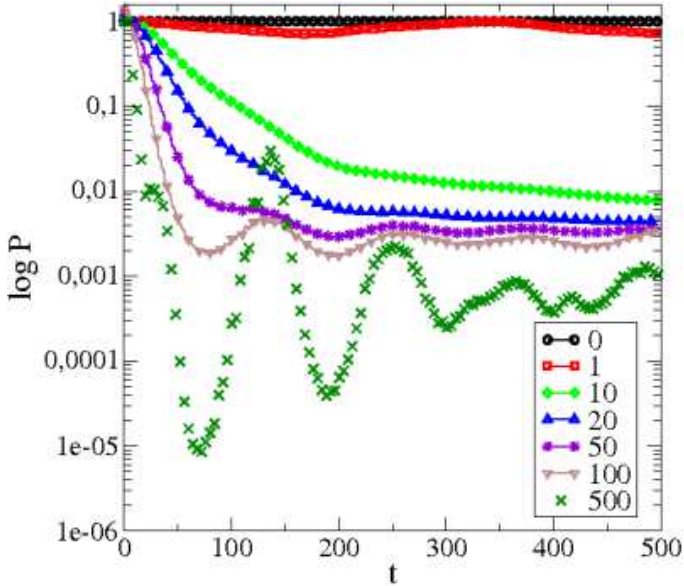}
  \caption{The purity for a bath with $N=1,10,20,50,100,500$ HOs.}
  \label{purity_ho}
 \end{center}
\end{figure}

In Fig.~\ref{purity_ho} we plot the time evolution of the
purity, Eq. (\ref{purity}), for the same parameters shown in 
Fig.~\ref{endecay_ho}.  For $N=0$ (no bath) the 
purity keeps constant to $1$. For $N=1$ the purity diminishes 
and increases a small amount in time, but no total decoherence 
is observed. For $N=10$ we have one exponential decay 
with decay exponent $\sim 0.024$ for times $t\lesssim67$ and 
another exponential decay with exponent $\sim 0.019$ for
times $75\lesssim t\lesssim 190$. For later time the purity 
decay obeys a 
power-law decay, as for the energy decay. In this case the 
total decoherence is observed for $t_{dec}\sim 140$ when the purity 
crosses the long-dashed straight line marking the value 
$P=0.067$. For the $N=20$ situation one can again notice three
decay regimes: two exponential decays with different exponents  
$0.047, 0.018$ (but they occur for the {\it same} time 
intervals as in $N=10$) and a power-law decay. The total decoherence 
is reached for $t_{dec}\sim 67$. For the purity, the fluctuating 
behavior for long times starts with $N=50$. By increasing $N$ the 
fluctuation's amplitude decreases.

Now, we explain the wavy pattern noticed for long times region
$t\gtrsim 100$. This phenomenon becomes more visible in the semi-log
plot for higher values of $N$ and it 
is a combination of two facts which act together. The first one 
is that the difference between the ground state energy level 
and the second energy level increases by increasing $N$, thus 
the ground state level becomes more and more predominant, 
compare Figures \ref{en_lev1} and \ref{en_lev10}. The second 
one is that the contribution of the ground state level to 
the energy exchange with the bath increases by increasing $N$.
For low values of $N$ the ground state energy is almost constant, 
while for $N \ge 50$ the ground state level gives part of the 
energy to the bath and gain it back at a later time, creating 
this fluctuating behavior. The minima of such fluctuations, occurs
close to times $t=60,190,310,\ldots$, which are very close to the 
times where a qualitative change in the energy and purity decay is 
observed in Figs.~\ref{endecay_ho} and \ref{purity_ho}.

\begin{figure}[ht]
 \begin{center}
  \includegraphics[clip=,angle=0,width=9cm]{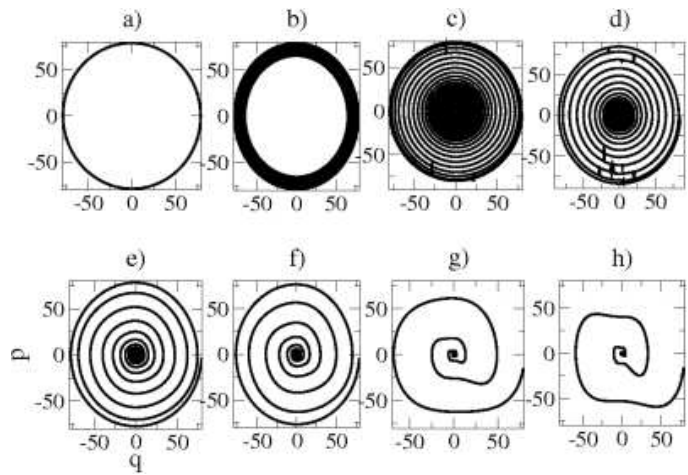}
  \caption{Phase space dynamics for $N=0,1,10,20,50,100,500,1000$ (from 
    left to right and top to bottom).}
  \label{phasespace_ho} 
 \end{center}
\end{figure}

Figure \ref{phasespace_ho} shows the phase space dynamics for one
realization of the bath and for different values of
$N=0,1,10,20,50,100,500,1000$. For $N=1$ the evolution occurs 
over an ellipsis with a small width. When $N$ increases to 
$10,20$ this width increases. It appears because the particle 
continuously loses some energy to the bath, but regain it later.
See Fig.~\ref{phasespace_ho}(c) for $N=10$ the particle motion
just for initial times. At $t=0$ the particle starts on the 
outer side of the ellipsis. As times goes on, part of its energy 
is transferred to the bath and it moves towards the inner side 
of the ellipsis. Since there is a continuous energy exchange 
between system and bath, the particles moves continuously 
between the
inner and outer side of the ellipsis. As $N$ increases
to
$100,500$ the strong dissipative character of the system appears 
clearly. Particle moves towards the origin in phase phase, 
and
stays there for all integrated times.

Now we turn our attention to other types of distributions, we consider the 
cases for which the spectral density has the form
$J(\omega)=\omega^s$, with 
$0 \le s \le 2$ \cite{r9}. If $s=1$ the
quadratic frequency distribution of 
the ohmic bath is recovered.
For $s \le 1$ we have the sub-ohmic bath, where 
lower frequencies
in the chosen domain have a bigger contribution when $N$ 
is sufficiently large. For $s \ge 1$ we have the super-ohmic case,
where 
higher frequencies appear more frequently.

In Fig.~\ref{expfits_ho} we plot the $\alpha$ exponents of the
short time 
exponential decay for all the three types of baths:
ohmic ($s=1.0$), 
sub-ohmic ($s=0.1$), and super-ohmic ($s=1.9$).
One can notice very small 
differences from one type of the bath to
another. For a better visualization 
we plot the results in a
log-log plot. One can clearly notice two distinct 
behaviors, shown
in the figure by continuous lines. For low values of $N$ 
the
exponent follows a power-law with the exponent equals to $0.749$
while 
for high values of $N$ the best fit is another power-law
function with the 
exponent equals to $0.362$.

In Figure \ref{powerlawfits_ho} we plot the $\beta$ power-law
exponents of 
the long times energy decay for ohmic, sub-ohmic, and
super-ohmic baths. As 
before, there were noticed small differences
in the exponents of different 
bath's type. We notice a fast
decrease of $\beta$ when $N$ increases. For 
large $N$, i.e.
$N>500$, it is difficult to say that the energy decay 
follows a
power-law function, due to the strong fluctuating behavior.

\section{Morse potential}\label{morse}

The Morse potential \cite{r7,r_mp,r10} is given by

\begin{equation}\label{M_pot}
  V(r)=D_e [1-e^{-a(r-r_e)}]^2,
\end{equation}

where $r$ is the distance between atoms, $r_e$ is the equilibrium
bound distance, $D_e$ is the well depth, $a=\sqrt{\frac{K_e}{2
D_e}}$ is the depth. $K_e$ is the force constant at the minimum of
the well.

\begin{figure}[ht]
\begin{center}
\includegraphics[clip=,angle=0,width=9cm]{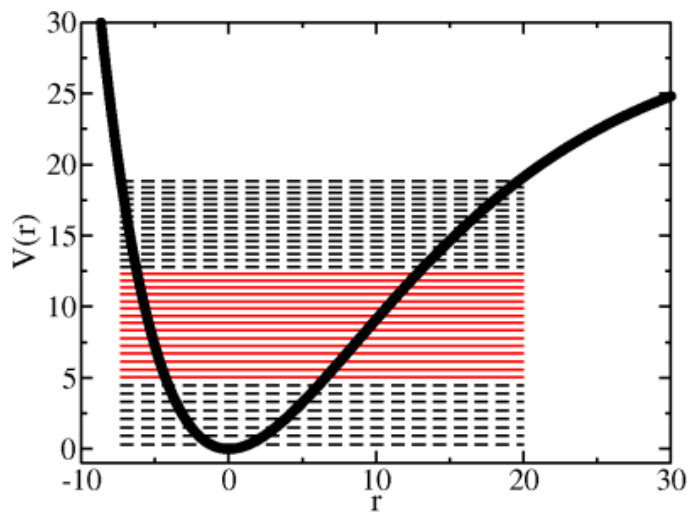}
\caption{Morse potential for $D_e=30, a=0.08, r_e=0$. With
horizontal lines are depicted the energy levels in the $r$ interval
$[-7.4,20]$, with continuous lines are shown the middle energies
levels used in our simulations (see the text for details).}
\label{m_p}
\end{center}\end{figure}

The energy levels $\epsilon_n$ and the position elements $r_{nm}$
were determined by using the numerov method \cite{r_num}. The
results for the energy levels were comparable with the analytical
form \cite{r7}:

\begin{equation}\label{en_mp}
 \epsilon_n = \hbar a \sqrt {\frac{2 D_e}{m}} (n+\frac{1}{2})-
\frac{[\hbar a \sqrt {\frac{2 D_e}{m}} (n+\frac{1}{2})]^2}{4 D_e}.
\end{equation}

Figure \ref{m_p} shows in thick continuous line the
Morse potential, given by Eq. (\ref{M_pot}), with the following
values of the constants \cite{r_mp}: $D_e=30, a=0.08, r_e=0$. With
horizontal lines are depicted all the $38$ energy levels, as
calculated from the numerov method. For the simulations we used as
initial condition for the coefficients $c_n$ a Gaussian wave
packet form $ \phi(r) = C \exp {- \frac{ (r -16)^2 }{\sigma^2} }$,
where $\sigma=3$ and $C$ is the normalization constant. Thus, we
will have a wave packet with the peak at the energy level $16$,
placed in the middle of the energy spectrum, corresponding to the
values $5.03$ to $12.32$, depicted by continuous horizontal lines
in Figure \ref{m_p}. Here, the coefficients $c_n$ have the real Gaussian form
defined above and they fulfill the equation $ \sum_{n=9}^{23} |c_n|^2 (0) = 1$.

\begin{figure}[ht]
 \begin{center}
  \includegraphics[clip=,angle=0,width=9cm]{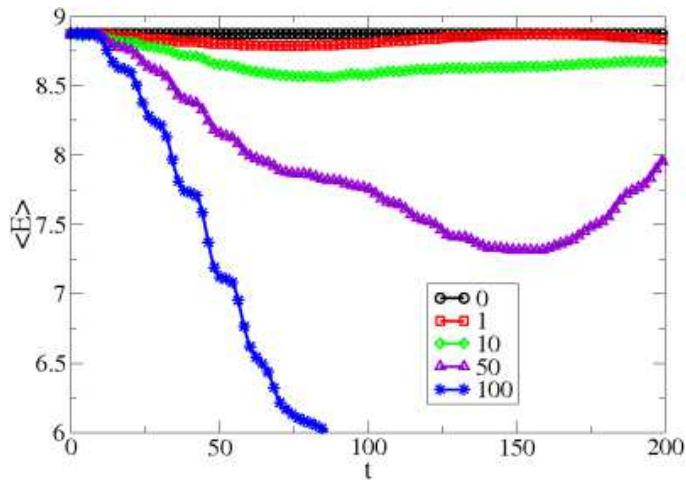}
  \caption{The energy decay, Eq. (\ref{energy}), for Morse potential. 
      The number of HOs in the ohmic bath is varied: $0,1,10,50,100$.}
  \label{endecay_ohm}
 \end{center}
\end{figure}

The energy decay, Eq.(\ref{energy}), for a bath having $N=0, 1, 10, 50,$ 
and $100$ HOs is plotted in Fig.~\ref{endecay_ohm}. It was considered 
that all of them have the same mass $m=1$ and experience the 
same coupling strength with the system, $g_{\lambda}=g=0.001$, which is
one order of magnitude smaller when compared to the HO case considered
before. Here, the 
frequencies of the oscillators
follow a quadratic distribution in the 
interval $[1.1,2.1]$, and
for each bath, once chosen the frequencies, are 
kept constant. For
each case, namely a bath having $N=1, 10, 50,$ and $100$ 
harmonic
oscillators, we will vary only $x_{\lambda}$ and $y_{\lambda}$,
which 
are real distributed numbers with zero mean and
deviation one. For 
all the non-zero cases ($N \neq 0$) we took
$500$ different initial 
conditions $(x_{\lambda},y_{\lambda})$, except the case of $N=100$, where due 
to the high CPU time we average over $200$ initial conditions. For the 
simulations we consider all the energy levels
of the system, described by a wave packet of $38$ energy levels,
$\epsilon_n$, corresponding to $38$ wavefunctions, $\phi_n$ (and
implicitly the $c_n$ 
coefficients). 

It can be easily seen that in the case of no bath $N=0$ the total energy 
is  conserved in time. When we have the case $N=1$, in the chosen time 
interval, the total energy decays slowly but returns to the system at 
$t\sim 150$. For $N\gtrsim 10$ HOs the energy decays exponential in time, 
with some fluctuations, due to the stochastic average. The power-law
decay observed in the HO case for later times is not seen here since the 
coupling to the bath is much smaller and decays occur much slower. In fact
the purity (not shown) also decays much slower when compared to the HO case.
\begin{figure}[ht]
 \begin{center}
  \includegraphics[clip=,angle=0,width=9cm]{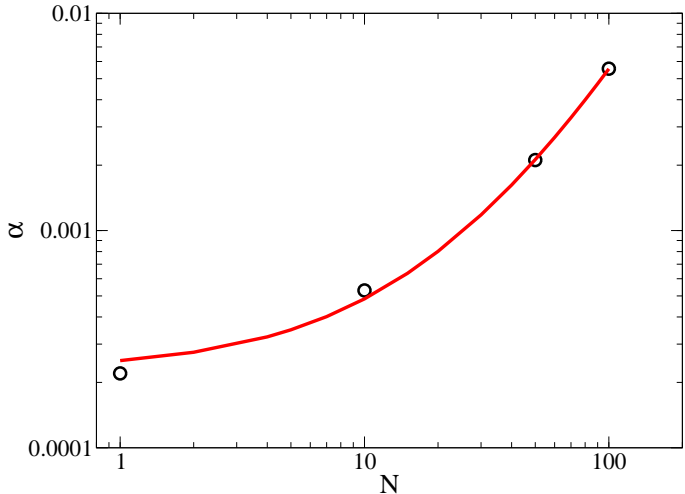}  
  \caption{The exponent $\alpha$ of the energy decay as a function of $N$. }
  \label{expfits_morse}
\end{center}
\end{figure}
The exponential energy decays can we summarized in Fig.~\ref{expfits_morse}, 
where the values of the exponential coefficient 
$\alpha$ of the energy decay are plotted as a function of the number of 
the oscillators of the bath, $N$. For a better visualization the results 
are shown in a log-log plot. The
situation is different than the power-law 
fits encountered for the harmonic potential. Here, we found that the best 
fit is a quadratic function given by $10^{-5} [0.0307 N^2 + 2.25 N +22.9]$.
Thus exponential decays increase with $N$.

\section{Conclusions}
\label{conclusions}
In this paper we use the stochastic Schr\"{o}dinger equation for zero 
temperature to study the energy decay for a system particle, situated in 
a harmonic potential and in a Morse potential. The system is coupled
to a bath composed of a finite number $N$ of uncoupled harmonic 
oscillators. For the discussions we chose, in the limit $N\to\infty$, 
an ohmic bath distribution, but we also study the cases of sub-ohmic 
and super-ohmic baths. 

In the case of a harmonic potential, with coupling intensity to the 
bath $\gamma=0.01$, it was observed that for very small numbers $N$
the energy is exchanged back and forth into the bath. For intermediate 
values of $N$ around $10\le N\le20$, the time average energy starts to 
decay, transferring partially its energy to the bath and we do not see 
any energy regain for the integrated 
times. For these values of $N$ we observed two exponential energy decays 
(with different exponents) for small times and a power-law decay for 
large times. For relatively higher values of $N$ ($\gtrsim 50$) a strong 
exponential decay was observed and no power-laws.
Therefore, non-Markovian dynamics is expected for intermediate values of $N$ 
($10\lesssim N\lesssim 20$) and a Markovian motion for larger values of 
$N$. The exponents for the exponential decays increase with $N$ obeying
two power-laws.
We analyzed also the purity of the system which gives informations about the
decoherence process. Essentially decoherence occurred in all cases for 
short times, where exponential decays were observed for both, the energy and
purity. For a system situated in a Morse potential, with coupling intensity
to the bath $\gamma=0.001$,  the high computational effort forced us to 
restrict our simulations to shorter time evolutions. Due to such small 
couplings, the energy and purity decays occur much slower when compared to 
the harmonic case and we just observed the exponential decays. Different 
from the harmonic case, the exponents of the exponential decay increase 
quadratically with $N$.

One of the main goals of this work is to show general features which 
occurs in systems coupled to an increasing number of bath constituents.
Therefore, in addition to 
the above main general results, we would like to conclude with some relevant 
characteristics observed in the simulations but not mentioned along the text: 
The energy decay rate increases with increasing coupling strength between 
system and bath, but the time for energy regain is independent of the coupling.
The energy regain time depends on the mean bath frequency and on $N$. For 
larger $N$ results are totally independent on the numerical frequency 
generator. For smaller values $1\le N\le 50$, the {\it values} of exponential
and power-law exponents for the energy decay may change for different 
frequencies generator, but the overall qualitative dynamics is equivalent.

\section* {Acknowledgements}
MG and MWB acknowledge the financial support of CNPq and FINEP (under the project CTINFRA). They also thank RM Angelo for discussions.

\addcontentsline{toc}{chapter}{Bibliography}
\end{document}